# Cancer progression as a learning process

Aseel Shomar[1,2], Omri Barak[3,2] and Naama Brenner[1,2]

[1]*Dept. of Chemical Engineering*, [2]*Network Biology Research Lab*, [3]*Rappaport Faculty of Medicine*

*Technion – Israel Institute of Technology, Haifa, 32000 Israel*

**Summary**

Drug resistance and metastasis - the major complications in cancer - both entail adaptation of cancer cells to stress, whether a drug or a lethal new environment. Intriguingly, these adaptive processes share similar features that cannot be explained by a pure Darwinian scheme, including dormancy, increased heterogeneity, and stress-induced plasticity. Here, we propose that learning theory offers a framework to explain these features and may shed light on these two intricate processes. In this framework, learning is performed at the single cell level, by stress-driven exploratory trial-and-error. Such a process is not contingent on pre-existing pathways but on a random search for a state that diminishes the stress. We review underlying mechanisms that may support this search, and show by using a learning model that such exploratory adaptation is feasible in a high dimensional system as the cell. At the population level, we view the tissue as a network of exploring agents that communicate and restrain cancer formation in health. In this view, disease results from the breakdown of homeostasis between cellular exploratory drive and tissue homeostasis.

**Introduction**

Cancer progression is traditionally viewed as the outcome of the accumulation of random genetic mutations and the selection of cells harboring mutations that confer them a growth advantage under certain conditions (Garraway and Lander, 2013; Nowel, P.C., 1976; Vogelstein et al., 2013). This Darwinian view is intuitively appealing and provided a powerful framework for cancer research since it reduced cancer to the molecular level of the DNA. Indeed, studies, that identified genes in which mutations are causally related to cancer, emerged at a brisk tempo, especially after the Cancer Genome Atlas and the International Cancer Genome Consortium were launched by the end of the 2000s (Martincorena and Campbell, 2015; The International Cancer Genome Consortium, 2010; Weinstein et al., 2013; Zhang et al., 2011).

This reductionist approach motivated developing drugs that target single molecular abnormalities or cancer pathways (Zugazagoitia et al., 2016). Although these drugs have achieved good clinical results, they only increased the overall survival by an average of nearly 3 months (Salas-Vega et al., 2017). A major reason for this is that malignant cells manage to adapt to the drug by bypassing its target, rendering the tumor resistant. In fact, resistance is the norm for all the drugs that target specific molecules (Vasan et al., 2019).

Another factor contributing to poor clinical outcome is metastasis – the process of migration and colonization of distant organs by cells from a primary tumor. Metastasis is the major culprit of cancer-associated mortality accounting for almost 90% of deaths (Cheung and Ewald, 2016; Lambert et al., 2017). These two avenues of disease progression and mortality – resistance and metastasis – are distinct biological processes. However, they seem to be remarkably analogous in several fundamental properties. In particular, accumulating evidence suggests that both drug resistance and colonization of cancer cells in secondary organs exhibit aspects that do not exclusively follow a simple Darwinian scheme of mutations and selection (Welch and Hurst, 2019). Despite many years of research, a clear genetic signature of mutations that are associated with metastasis has not been found (Lambert et al., 2017; Vogelstein et al., 2013). Such a signature was found in some cancers at the initiation stage and was therefore expected and sought also in the context of metastasis. Resistance, in parallel, can arise from purely non-genetic mechanisms (Bell and Gilan, 2020; Marine et al., 2020; Pisco and Huang, 2015).

A cytotoxic treatment or a hostile environment in a secondary organ are both sources of stress to which cancer cells should adapt in order to survive. Plethora of findings showed that such a stress can actively induce an adaptive response and lead to drug resistance and the formation of metastasis (Pisco and Huang, 2015; Welch and Hurst, 2019). Such adaptation is supported by global epigenetic changes that enhance the plasticity of the cells. Various terms have appeared in the literature to describe these phenomena: 'alternative pathways', 'acquired', 'adaptive' or 'induced' resistance (Kim et al., 2018; Maynard et al., 2020; Pisco and Huang, 2015; Shaffer et al., 2017; Stewart et al., 2020). Nevertheless, a comprehensive theory and deeper understanding of these terms is still lacking; such an understanding is crucial to pave the way to new therapeutic avenues.

In this Perspective, we propose that learning theory offers a framework that may shed some light on the two complex processes of drug resistance and metastasis. Learning is the fundamental ability of a system to modify itself with relation to its environment and thus acquire novel functionality. Although most commonly attributed to the brain, we support the view that learning is not an exclusive function of neural networks, but rather a general property of plastic high-dimensional systems coupled to their environment (Baluška and Levin, 2016). Experiments on yeast cells have revealed that cellular networks can provide a substrate for learning and adaptation in the face of unforeseen challenges to which no pre-programmed response is available (Braun, 2015). This experimental paradigm was based on artificially rewiring the genetic regulation network of yeast cells. Exposing these cells to an environment where expression of the rewired gene is essential while its natural regulation is compromised, confronts them with regulatory challenging and stressful conditions for which no pre-programmed response is available. Without dedicated sensory information or pre-existing regulatory pathways, the yeast cells adapt through a trial-and-error process that can be described as a primitive form of learning by the gene regulatory network (Schreier et al., 2017).

Here, we propose that a similar type of adaptation is implicated in metastasis and resistance, where cancer cells aim to achieve stress reduction. Stressful situations, such as drugs and lethal environments, induce plasticity that can initiate an exploratory process by which new states of the cellular networks – regulatory genetic network, metabolic and signaling networks – can be sampled. As a more adaptive and less stressful state is reached, the drive for additional sampling of new states decreases. The novelty of our approach is to highlight a mode of adaptation which is not contingent on pre-programmed pathways but rather on the ability of the cells to explore and contrive new adaptive states, and to establish the deep connection of such adaptation with a learning process.

We combine these insights from learning theory with the tissue population level of organization. While each individual cell can be viewed as a learning system in itself and might explore to find an adaptive state, it is also part of a coupled cell population in a tissue environment and interacts with neighboring cells. Thus, its propensity to explore might be constrained at the population-level by a collective force that drives the cells to a local synchronous state.

To make our framework more concrete, we address two corollary questions: First, can one identify empirical signatures of exploratory dynamics in the processes of resistance and metastasis? We will delve into corroborative findings that highlight prominent features of exploration and tissue constraints in cancer progression. Second, how is exploration feasible as an effective converging mechanism in a complex high dimensional system as the cell? We will discuss the feasibility of such a mechanism based on recent modeling work done in our lab.

**a** Dormancy

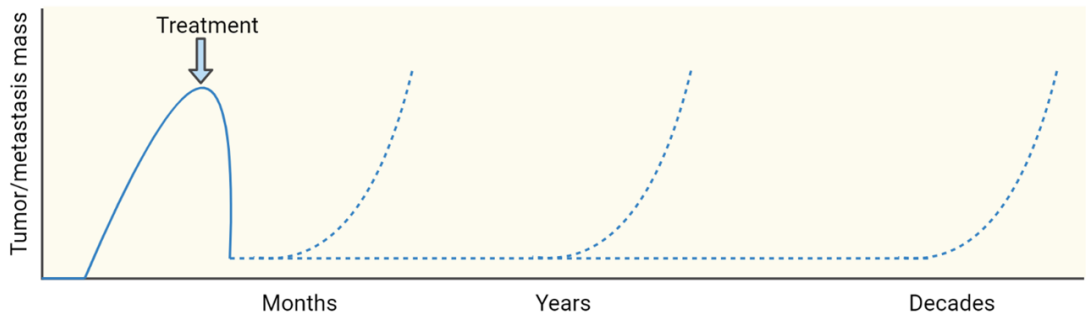

**b** Induced heterogeneity

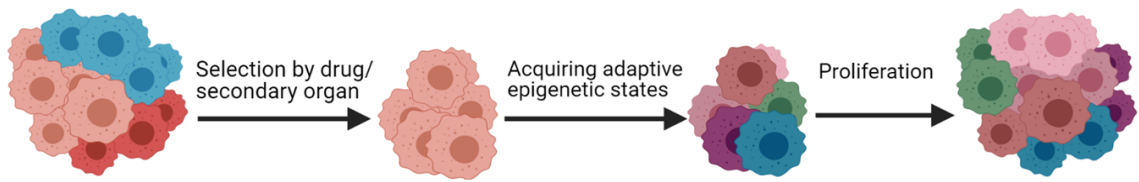

**c** Stress-induced stemness

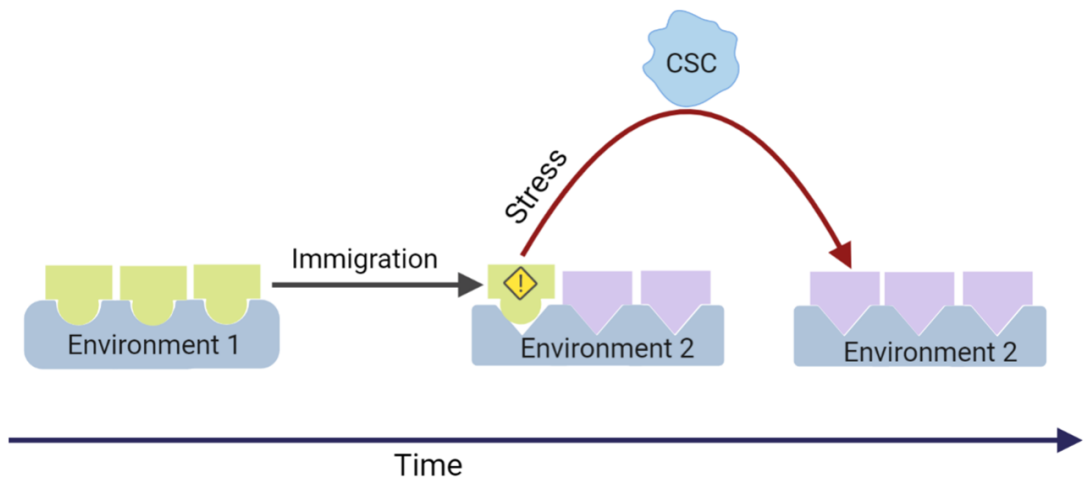

**Figure 1. The fingerprints of exploratory adaptation in resistance and metastasis. (a)** Dormancy: stressed cancer cells often enter a non-proliferative state that might be crucial for them to contrive adaptive states. **(b)** Induced heterogeneity: despite the selective pressure of the treatment or the secondary organ, resistant cells and metastases exhibit high heterogeneity. This is concordant with exploratory adaptation as it yields multiple solutions to the same problem. **(c)** Stress-induced stemness: The transition to a stem cell state is enhanced when cancer cells are exposed to stress such as a new environment. This transition provides cells with high plasticity that enables them to search for a new adaptive state. Created with BioRender.com.

**The fingerprints of exploratory adaptation in resistance and metastasis**

*Long and variable timescales in Dormancy*

Dormancy[1] is a prominent feature of both resistance and metastasis, uncharacteristic of a pure Darwinian scheme. Cancer cells that arrive at distant organs can remain in a quiescent non-proliferative state for a highly variable time period that extends up to decades (Fig. 1a) (Chaffer and Weinberg, 2011; Gupta and Massagué, 2006; Lambert et al., 2017). The clinical implication of this phenomenon is that dormant cells can cause a metastatic relapse even years after radical dissection of the primary tumor.

Similarly, in response to cytotoxic treatments, small subpopulations of cells can survive by initially entering a drug-tolerant state that displays little to no proliferation (Bell and Gilan, 2020; De Angelis et al., 2019; Marine et al., 2020; Ramirez et al., 2016; Sharma et al., 2010). These cells are genetically indistinguishable from the bulk tumor population, and they regain drug sensitivity after its withdrawal. Crucially, a fraction of these cells can gain the ability to proliferate in drug after a long-term treatment (weeks to months). Thus, the quiescent subpopulation provides an adaptive reservoir for the acquisition of resistance mechanisms (Bell and Gilan, 2020).

In the absence of overt proliferation, genetic changes are less likely to occur and to be selected for (Giancotti, 2013). This adds to the arguments suggesting that adaptation to drugs or new environment may not occur through the selection of accumulating resistant genotypes. What exactly happens during the dormancy period is not well understood; it has been suggested that stress signaling is involved, as well as the emergence of stem-cell-like properties, which will be discussed in more detail below (Giancotti, 2013). One possibility is that dormancy time is utilized for an active search process in a large space of possible phenotypic configurations towards an epigenetic adaptive state. Once the cells manage to contrive an adaptive phenotype, they exit this state and reactivate growth. The yeast rewiring experiments provide an opportunity to glimpse into the single cell dynamics during this latent phase. Woronoff et al (Woronoff et al., 2020) utilized micro droplet techniques to reveal that different yeast cells in the latent phase consume sugar at varying rates. This rate was correlated with their adaptation success, suggesting that the energy invested in metabolic activity might be the cost of contriving adaptation mechanisms during this period.

---

[1] Several types of non-proliferative cell states have been characterized in the literature, for example dormant, quiescent, senescent; this includes the distinction between different states by identification of specific markers. From a physiological point of view, the crucial phenomenon is the appearance of long timescales of very slow, or complete absence, of cell growth and division. In our context these can be grouped into a non-proliferative phenotype.

In cancer cells, no such measurement is available. An indirect clue might be obtained by observing the state of cells at the end of the latent phase. Specifically, does passage through a dormant state force cancer cells into a single genetic/epigenetic state, or can multiple resistant phenotypes eventually arise? Exploratory adaptation supports the latter scenario. The nature and extent of heterogeneity will be considered next.

*Single cell heterogeneity*

Both resistant and metastatic cells are highly heterogeneous (Fig. 1b) (Gupta and Massagué, 2006; Klein, 2013; Lawson et al., 2018; Scott et al., 2012; Vogelstein et al., 2013). The heterogeneity of metastatic lesions is especially intriguing in light of the fact that metastasis is an extremely inefficient process. The bottleneck of metastasis is colonization in distant organs where cells need to adapt to a new deadly environment – even the most congenial environment is still very hostile for cancer cells. While 80% of tumor cells that are injected into the circulation manage to survive and extravasate, less than 0.02% form macro-metastases (Brabletz, 2012; Giancotti, 2013; Luzzi et al., 1998; Massagué and Obenauf, 2016). In this process, the miniature fraction of cells that manages to survive the new pernicious environment produces high intra-metastatic heterogeneity. This is somewhat surprising when compared to the effect of bottlenecks in ecology, where the signature of species migration is a dramatic temporary decrease in population heterogeneity (Amos and Harwood, 1998). Such a decrease is expected in any scenario that is based on selection at the population level, regardless of the mechanisms underlying the generation of variability. Thus, the heterogeneity of cancer cells following colonization of a new tissue, consistently with the temporal features discussed above, suggest that the process does not rely solely on the selection of fit cells but contains an element of intracellular dynamics.

Concordantly, recent works showed a globally increased intra-tumoral heterogeneity following treatment resistance (Jr et al., 2016; Stewart et al., 2020). This increased heterogeneity appears to be drug-induced, since tumors treated with a control vehicle exhibited less heterogeneity than resistant tumors. However, in contrast to the induction of pre-existing pathways dedicated to respond to a specific signal, here cells exposed to the same signals activate different pathways.

Taken together, these observations suggest that both metastasis and drug resistance include a crucial element of intracellular dynamics that results in highly variable outcomes. While both dormancy and heterogeneity can be considered indirect evidence of a highly plastic state of cancer cells, the next section highlights a more direct observation of plasticity.

*Stress-induced stem cell state*

The concept of a stem-cell represents an extremely high degree of cellular plasticity, with very little constraints from epigenetic barriers. Although once thought to be a starting point of a unidirectional path in development leading to cell differentiation, it is now known that this path is reversible (Sánchez Alvarado and Yamanaka, 2014). Non-specific induced resistance can be implemented by a stem-like cell state, with a larger potential to acquire new phenotypes that are not accessible to fully differentiated cells; indeed, stemness has been tied to resistance-enhancing plasticity in many studies (Adorno-Cruz et al., 2015; Chang, 2016; De Angelis et al., 2019; Doherty et al., 2016; Lytle et al., 2018; Pisco and Huang, 2015). Induction of stemness can occur through a non-proliferative state (De Angelis et al., 2019). For example, chemotherapy can initiate such a sequence of events that results in senescence-associated stemness (Milanovic et al., 2018).

Cancer stem cells are also pivotal players in metastasis, and were suggested to possess a tumor initiating potential and exhibit a higher degree of plasticity that enables them to adapt to the challenges posed by the new environment (Adorno-Cruz et al., 2015; Chang, 2016; Doherty et al., 2016; Lytle et al., 2018). In particular, stem-like plasticity properties can be induced by external signals that are linked to the Epithelial Mesenchymal transition (EMT), which is considered an early step in metastasis formation (Doherty et al., 2016). It is intriguing that the same noxious environment, that is supposed to kill the cells, enables them to acquire essential traits for survival. This dual role of the environment in metastasis is analogous to the role of the drug in resistance; in fact, metastasis can be viewed as resistance to the new environment. However, in contrast to drugs which target one pathway, survival in a new environment requires myriad general and tissue-specific adaptations.

For instance, cancer cells face a higher oxidative stress in the target organs; thus, cells that manage to produce antioxidants are more prone to survive (Piskounova et al., 2015). In addition, many organ-specific adaptations in the lung, bones, brain and liver were identified (Massagué and Obenauf, 2016; Obenauf and Massagué, 2015). In the brain, for example, astrocytes produce plasminogen activator which induces the production of plasmin that leads to cancer cell death. Thus, to survive under such conditions cancer cells need to produce serpins that are typically produced by neurons to shield them from plasminogen activator-mediated cell death (Valiente et al., 2014). In the liver, the survival of cancer cells is highly associated with their ability to consume creatinine and ATP to produce phosphocreatine which endows them with a survival advantage (Loo et al., 2015). Such fundamental phenotypic changes require an extremely plastic cell state, such as that provided by stem cells (see Fig. 1C).

**Is exploratory adaptation feasible?**

Taken together, the three features described in the preceding sections combine to suggest that single cells, in the context of metastasis and resistance, can enter a highly plastic state in order to explore intracellular configurations that may lead to adaptive behavior. Motivated by these observations, we proceed to examine the feasibility of exploratory adaptation.

To this end, two questions should be addressed – pertaining to the parts and to the whole: First, what are the molecular building blocks that could lead to trial-and-error learning? Second, does it all add up? Namely, can the process of exploratory adaptation converge within the context of single cell networks?

**What molecular mechanisms support trial-and-error learning?**

When drug resistance develops, even if selection of pre-adapted sub-populations is involved, cells still need to undergo epigenetic reprogramming to acquire resistance (Hong et al., 2019; Kim et al., 2018; Maynard et al., 2020; Shaffer et al., 2017; Sharma et al., 2010). Interestingly the acquired resistance is neither drug specific nor pathway specific (Sharma et al., 2010), but rather entails global broad epigenetic changes. The coupling between cellular stress and mechanisms that can promote such changes is a central building block in our suggested framework (Braun, 2015; Soen et al., 2015). Stress can drive temporary plasticity, in turn driving global epigenetic reprogramming, and eventually allowing the acquisition of novel adaptive phenotypes. A recent study has shown that tumors consistently contain a fraction of cells in a stress-like state (Baron et al., 2020). Intriguingly, these cells are more efficient in seeding new tumors and hold drug-resistant properties that can be induced by heat shock.

An important mechanism that can increase cellular plasticity and enable trial-and-error exploration is chromatin remodeling. Disruption of chromatin homeostasis can lower epigenetic landscape barriers, making large regions of the genome accessible to transcription factor (TF) binding, and expanding the attainable space of gene expression patterns (Fig. 2a) (Flavahan et al., 2017; Guo et al., 2019). Importantly, chromatin remodeling can be induced by non-genetic factors and provide transient and reversible coupling between stress and epigenetic plasticity. For instance, overexpression of the transcription factor Nfib led to increased accessibility to distal regulatory elements that promote pro-metastatic neural gene expression programs (Fig. 2a) (Denny et al., 2016). Chromatin remodeling was also found to dynamically mediate resistance in a subpopulation, in response to drug application (Sharma et al., 2010). Thus, chromatin remodeling emerges as a candidate mechanism for modulating the level of cellular plasticity in a transient and stress-sensitive manner. In support of this picture, yeast cell experiments showing exploratory adaptation following a rewiring perturbation, exhibited variation in the efficiency of adaptation in strains with mutations associated with chromatin remodeling (Freddolino et al., 2018).

Both stress responses and chromatin remodeling have a global effect on the cellular network. In addition, the local elements of the network are themselves plastic. Many proteins have alternative binding partners that in turn give rise to multiple binding patterns depending on context, each resulting in markedly different network configuration (Fig. 2b). For TFs, which mediate regulatory connections, this can induce a flexible network structure. Genome-wide binding assays in yeast have shown that TF binding patterns are context-dependent (Holland et al., 2019; Lee et al., 2002). In mammalian cells, it was demonstrated by computational analysis of time-varying single-cell data that edges in the regulatory networks are modulated during induced EMT (Krishnaswamy et al., 2018).

Alternative binding of transcription factors could be supported by the coexistence of multiple conformations (James and Tawfik, 2003). Intrinsically disordered protein regions, once thought to be a curious feature of a small number of proteins, are now acknowledged as a general property of many proteins, and most notably of TFs (Peng et al., 2015). Such disordered regions can give rise to a large number of folds and binding affinities (Fig 2b). While the role of random protein domain is still under intense study, recent results from plants implicate their connection with adaptation to stressful conditions (Liu et al., 2017). The existence of multiple alternative functional folds of TFs can potentially provide a powerful mechanism to confer plasticity to the genetic network and enable to explore different configurations due to rapid fluctuations between different conformations (Wright and Dyson, 2015). Other molecular mechanisms, such as post-transcriptional modifications, have also been suggested as an important characteristic of TFs that induces flexibility on gene regulation (Niklas et al., 2015).

From a systems-level perspective, such plasticity represents the ability of cellular networks to modify themselves, in analogy to neural networks in learning. The main mechanism thought to support learning in neural networks is synaptic plasticity, the ability of connections between neurons to remodel following signals and experience. Connection remodelling also underlies learning in practically all artificial network algorithms. The analogy with the ability of gene regulatory networks to modify their interactions, as supported by the above mechanisms, is straightforward and would thus endow the network with the ability to learn. This analogy motivated us to address the feasibility of convergence by borrowing concepts from learning models traditionally used to study neural networks.

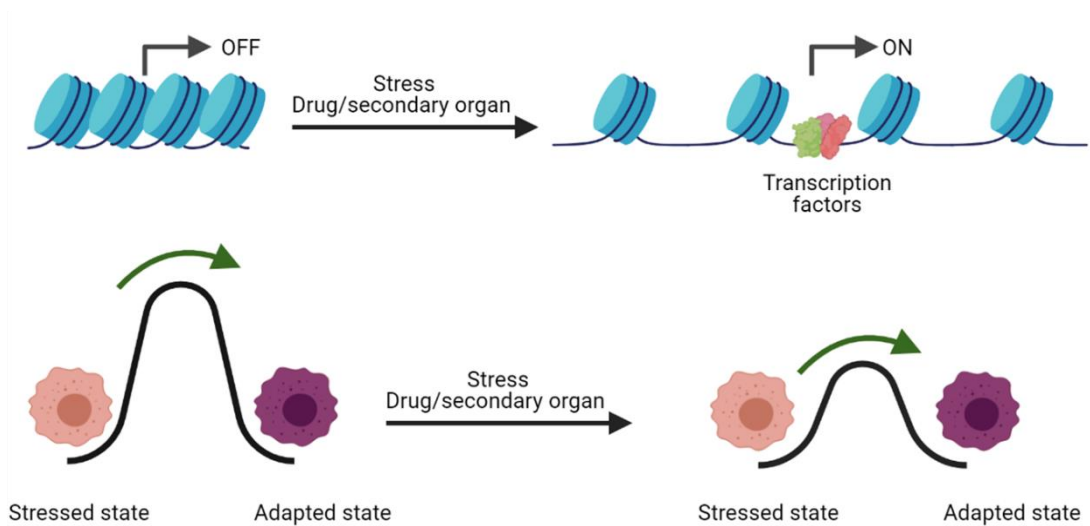

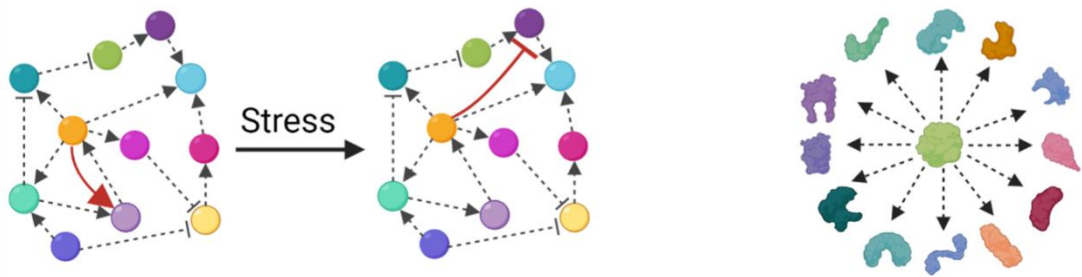

**Figure 2. Molecular mechanisms supporting trial and error learning. (a)** Chromatin remodeling: (top) stress induced by the secondary organ or the drug can reshape the chromatin landscape, making it more permissive and active. (bottom) This endows the cells with higher plasticity to explore alternative states by lowering the barriers of transition between them. **(b)** Dynamic regulatory network: (left) Regulatory networks are dynamic and transcription factors can even take contradictory roles depending on context. Arrows represent activating interactions and caps represent inhibitory interactions. (right) Intrinsically disordered proteins can confer plasticity by alternating between different conformations. Created with BioRender.com

**Can a high dimensional system converge under exploratory adaptation?**

In contrast to evolutionary trial-and-error dynamics, which takes place at the population level, exploratory adaptation involves trial-and-error at the level of the individual. Variation is created in a single system along time and selected by feedback from the environment. Such dynamics have been considered in several biological contexts, such as spindle assembly and bacterial chemotaxis (Fig. 3a-b) (Kirschner and Gerhart, 2005). However, these instances of exploration take place in three-dimensional space, and therefore, are feasible in terms of convergence success at finite time. For cancer cells to adapt by exploring gene expression, they face a different situation of a high-dimensional system (order of thousands of degrees of freedom) undergoing some random search in its vast configuration space (Fig. 3c). With no sophisticated means to compute an error from some target function, the cell has only its global stress level at its disposal to provide feedback on the exploration. Under what conditions can such a learning scheme converge successfully? Since this is a quantitative question, one naturally turns to mathematical modeling: a model that captures the essential features of exploratory adaptation in high-dimensional space can shed light on the conditions for its possibility of convergence.

We constructed such a model based on large random networks of interacting elements (Schreier et al., 2017). This is a popular modeling approach for gene regulation when general properties are of interest, such as number of fixed points, evolvability or canalization (Drossel, 2008; Kauffman, 1993; Li et al., 2013).

The model (Fig. 3d) describes a system of N interacting genes (N on the order of 1000), whose state (expression level) is represented by the N-dimensional vector $\boldsymbol{x} = (x_1, x_2 \ldots x_N)$. Their time evolution is determined by a nonlinear dynamic rule governed by a random interaction matrix $W$, in which the element $W_{ij}$ represents the strength of influence of gene $j$ on gene $i$. Regulatory plasticity is represented by the ability of the interactions $W_{ij}$ to change their strengths over time; in exploratory dynamics, these changes will be essentially random and their amplitude controlled by feedback. The critical ingredient of the model is to define the feedback and close the loop.

A given cellular phenotype can be realized by different expression patterns. Mathematically, this corresponds to a low-dimensional projection, which we take to be linear, with weights given by arbitrary numbers $\boldsymbol{b}$ :

$$y(t) = \boldsymbol{b} \cdot \boldsymbol{x}(t) \tag{2}$$

The challenge is presented to the system by the constraint of maintaining the phenotype in a range $y(t) \approx y^*$. This allows for multiple gene expression patterns to comply with the constraint in different microscopic implementations. If, however, the projection deviates outside an allowed range, a global cellular stress will emerge, $S(y - y^*)$, causing the system to initiate exploratory dynamics in the form of random changes in the interaction matrix $W$. The amplitude of these changes is dictated by cellular stress $S$: it relaxes if and when it reaches a stable state that relieves stress, $S \approx 0$ (Fig. 3b). In this way, a simple feedback loop couples internal exploratory dynamics to the suitability of the current configuration to relieve stress and to match the environmental demand.

One might imagine that such a simple algorithm will not allow convergence to an appropriate phenotype in a reasonable time for large networks, and indeed this is the case for homogeneous random matrices $W$, corresponding to identical and independent probability of connections between any two genes. Intriguingly, the main conclusion from the modeling work was that exploratory adaptation is sensitive to network structure; in particular, it is likely to converge for networks with outgoing hubs - a small number of nodes with disproportionately high connectivity (Fig. 3c). For example, in a scale-free network topology there are hubs in the tail of the distribution that are connected to a large fraction of the network (Schreier et al., 2017). However, scale-free networks are merely a mathematical tool to classify topologies; we have shown that a handful of outgoing genes in an arbitrary network are sufficient to induce the effect. In real gene networks, master regulators are well known genes that control up to hundreds of other genes (Cai et al., 2020). Such hubs are usually considered in the context of specific gene programs that they regulate, but our model proposes that they can also coordinate network plasticity and drive it more easily through an exploratory process to discover novel stable states. This new role is similar to the stabilizing effect of an external feedback that suppresses irregular network activity (Rivkind et al., 2020). Importantly, it is in line with recent discoveries on master regulators in the context of cancer reprogramming.

For instance, ZEB1 is a key player in EMT, where its most well-known function is the suppression of epithelial genes. Interestingly, under some conditions the flexible nature of this regulator is revealed and ZEB1 can turn into a transcriptional activator in aggressive cancer types (Lehmann et al., 2016). In fact, this is not a special property of ZEB1, but also exists for other transcription factors (Stemmler et al., 2019). Thus, master regulators can direct alternative expression patterns in different situations. In the absence of an external agent that directs the hubs which program to choose, our framework of exploratory adaptation suggests that the appropriate configuration emerges through stress-mediated feedback.

In conclusion, the feasibility of exploratory adaptation in structured random networks, and the finding that actual gene regulatory networks fulfill the key requirements of the model, suggest that this mechanism can be implemented in cells adapting to stressful conditions. In particular it can help explain the behavior of cancer progression in drug resistance and metastasis. Intriguingly, applying this model to experimental data, Celiku et al (Celiku et al., 2019) have shown that the adaptation of glioblastoma cells as they spread to diverse tumor microenvironments can be at least partly attributed to exploratory dynamics. Future work may use this framework and model to analyze experimental data in other systems.

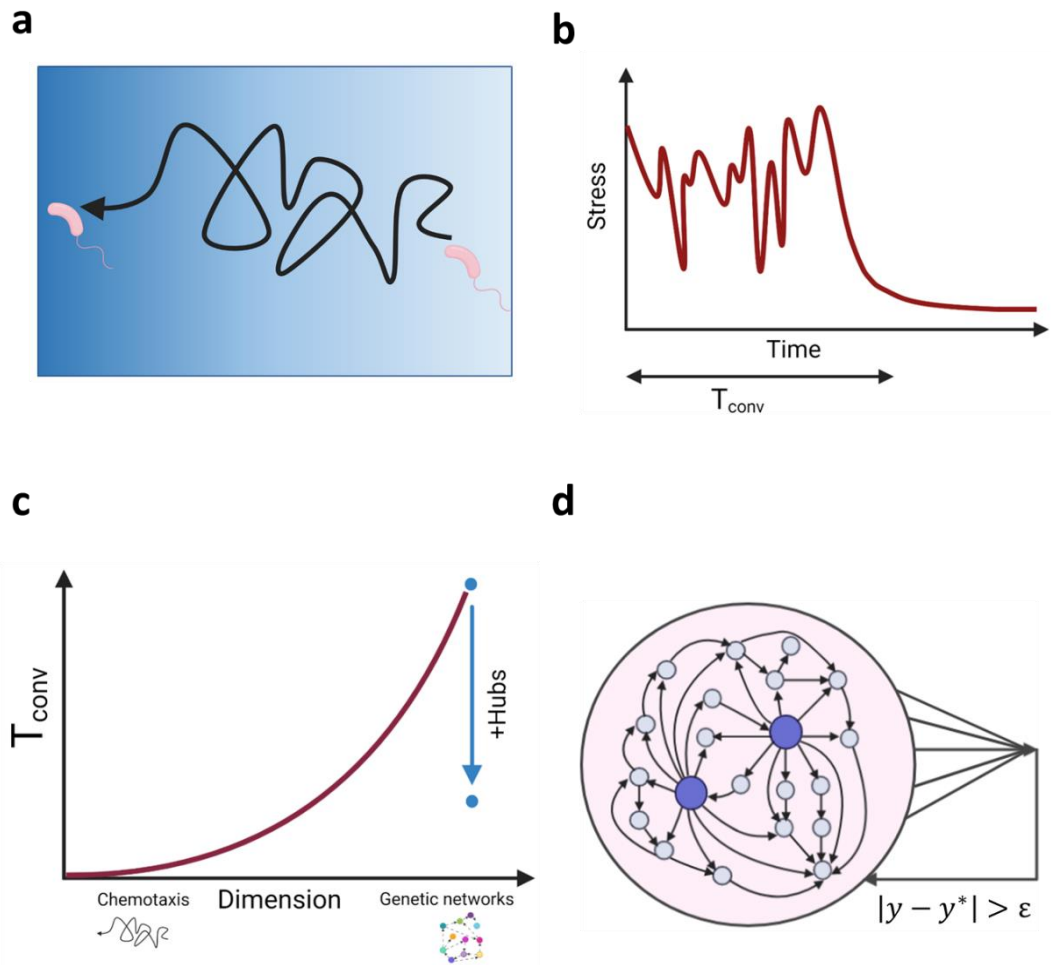

**Figure 3. Modeling exploratory adaptation. (a)** A low-dimensional example of exploratory adaptation is chemotaxis, where bacteria move toward a higher gradient of an attractant (blue gradient) by random walk. **(b)** When a system is exposed to stress, it aims to search a new state that relaxes the stress through a trial-and-error exploration. When the system reaches such a state, exploration stops. The time that takes the system to reach this state is called convergence time ($T_{conv}$). **(c)** Convergence time increases dramatically with the number of dimensions. In the case of a genetic network, each gene is a dimension. This makes it harder for networks performing exploratory adaptation to converge. Nevertheless, networks that harbor outgoing hubs- nodes with disproportionately high number of outgoing connections- can converge in a plausible time. **(d)** Cells exposed to stress perform exploratory adaptation in which the connections between the genes are modified by random walk. The strength of the random walk is dictated by the distance between the low dimensional state of the cell, $y$, and the constraint $y^*$. Outgoing hubs (big purple nodes) enable exploratory adaptation to converge. Created with BioRender.com

### The constraining force of the population

The capability of cells to explore for new configurations makes them vulnerable to fall by chance into cancerous configurations when they need to adapt to a stress. Without any constraining force, cancer would be the norm rather than the exception. We will argue below that this is not the case because the tissue constrains exploration and cancer progression. There has been an increasing surge of interest in the effect of cellular communication in cancer and many excellent reviews gather previous findings in this area (Bissell and Hines, 2011; Capp, 2005; Soto and Sonnenschein, 2011). Here, we tie these findings to our learning framework and emphasize the significance of cellular communication as a constraint on exploratory adaptation.

**Mutations and growth control in normal tissues**

Cancer is traditionally associated with the occurrence of specific patterns of mutations that are thought to drive transformation. It turns out that these mutations are not sufficient for transformation. A recent comprehensive RNA sequencing analysis detected thousands of somatic mutations across all human tissues and in almost all tested individuals, including mutations at cancer hotspots and other cancer genes (Yizhak et al., 2019). Greater numbers of mutations were observed in tissues that are more exposed to carcinogenic environmental factors, such as sun-exposed skin, esophagus mucosa and lung (Yizhak et al., 2019). Congruently, a recent analysis of sun-exposed skin tissues revealed a remarkably high burden of somatic mutations of averaged two to six mutations per megabase per cell, similar to that found in many cancers (Martincorena et al., 2015). The frequency of mutations that were previously identified as "driver mutations" in these tissues was surprisingly high. For instance, there were more NOTCH1 mutations in a sun-exposed skin biopsy than have been identified in more than 5000 cancers sequenced by The Cancer Genome Atlas (Martincorena et al., 2015). Remarkably, clones carrying driver genes expand in normal skin tissues; however, this growth stops early in the expansion, giving rise to limited sized clones (Martincorena et al., 2015). Some of the analyzed clones carried two to three driver mutations while having a totally normal skin phenotype. Similarly, sequencing studies of normal blood cells revealed signatures of somatic mutations broadly similar to blood cancer (Genovese et al., 2014; Jaiswal et al., 2014). These findings raise the question: how is homeostasis maintained in normal tissues despite the high burden of cancer driver mutations? What prevents the appearance of cancer phenotype in these tissues?

One widely-accepted explanation is that driver mutations trigger oncogene-induced senescence (OIS), arresting the proliferation of cells before the accumulation of additional mutations (Bennett, 2003; Huang et al., 2017; Kaplon et al., 2014; Michaloglou et al., 2005; Serrano et al., 1997; Wajapeyee et al., 2008). OIS is usually depicted as a cell-autonomous stress response, namely expressing an oncogene in a cell leads to stress which induces a growth arrest in that cell. However, a recent work by Ruiz-Vega at al. (Ruiz-Vega et al., 2020), using a mouse model of BRAF-driven nevus formation, showed that this is not necessarily the case. BRAF mutation is the most common driver mutation in melanoma (Davies et al., 2002), yet it is present in 89% of benign nevi (pigmented 'moles') (Pollock et al., 2003). In this work, no evidence supported the senescence of nevus cells, either compared with other skin cells or other melanocytes. Moreover, nevus size distribution could not be fit by any simple cell-autonomous model of growth arrest yet, were easily fit by models of collective feedback between the cells. This emphasizes the significance of the tissue level of organization, in particular cell-cell interactions, in maintaining homeostasis and constraining cancer development.

**For better or worse: micro-environment controls cell fate**
Intriguingly, the effect of cellular communication is so dominant that it not only stops the development of a cancerous phenotype, but can actually turn around cell fate. The classic work of Beatrice Mintz and Karl Illmenesse (Mintz and Illmensee, 1975) showed that placing teratocarcinoma (undifferentiated embryonic carcinoma cells) in a blastocyst gave rise to perfectly normal and tumor free offspring that displayed many traits of the parental tumor cells. Similarly, transplantation of nuclei from malignant cells in enucleated oocytes gave rise to stem cells that were able to produce mice (Hochedlinger, 2004).

Numerous works provided additional evidence for the ability of the micro-environment to suppress tumor growth and induce differentiation to a variety of functional tissues. In their seminal work, the laboratories of Ole Petersen and Mina Bissel showed that breast cancer cells revert to nearly normal phenotype when cultured in three-dimensional culture that mimics the normal breast tissue (Howlett et al., 1994). The genome of the reverted cells was shown to be similar to mutated and malignant cells grown in two-dimensional cultures (Rizki et al., 2008; Weaver et al., 1995). Similarly, highly malignant melanoma cells injected into Zebra fish embryos (Kasemeier-Kulesa et al., 2008), mammary carcinoma cells recombined with normal mammary gland stroma (Maffini et al., 2005), and liver cancer cells injected into normal liver (McCullough et al., 1997) are all additional examples for the ability of the microenvironment to normalize cancer cells. More recently, it was shown that converting invasive breast cancer cells into adipocytes by treating them with appropriate cues inhibits cancer metastasis (Ishay-Ronen et al., 2019). Unequivocally, all these findings point to the fact that the environment can play a crucial role in redirecting the phenotype of cancer cells and determining whether cancer is contained or spreads. They also demonstrate the high plasticity of cancer cells which can be exploited to adapt to dynamic changes, triggered by external signals.

Unfortunately, the stroma (fibroblasts, vasculature, immune cells and interstitial ECM) that suppresses the growth of tumors, can induce tumorigenesis when it undergoes detrimental changes. Maffini et al. (Maffini et al., 2004) showed that combining a carcinogen-exposed mammary stroma with vehicle-exposed mammary epithelium resulted in neoplasm. The reverse combination did not. Similar results were obtained from a normal mammary cell line with an irradiated stroma (Mh and Sa, 2000), and a normal prostate cell line and fibroblasts derived from prostate cancer(Barclay et al., 2005). These findings indicate that carcinogenesis can be the result of an abnormal interaction between the stroma and epithelial cells. Moreover, such interaction can support tumor progression and induce resistance (Chan et al., 2019; Shaked, 2019).

These arguments suggest a balance between the exploratory drive of individual cells under stress, and the constraining effect of interaction with the environment in a healthy tissue. The emerging picture suggests a view of cancer which goes beyond the single-cell and places the disease at the level of the cell-tissue interface. In this framework cellular communication is a double-edged sword (Bissell and Hines, 2011). It can prevent transformation of normal cells harboring mutations and normalize cancer cells despite their mutations, but also induce tumorigenesis when aberrant. As Smithers stated it: "Cancer is no more a disease of cells than a traffic jam is a disease of cars. A lifetime study of the internal combustion engine would not help anyone to understand our traffic problems"(Smithers, 1962).

**Conclusion**

Taken together, we propose that single cells have the capability to learn novel phenotypes by utilizing their internal plasticity and under the direction of global cellular stress. This proposition is based on analogies between features of cellular adaptation to stress and learning in neural networks. We view cells in a tissue as a system of coupled explorers or learners. The state of the entire system is determined by an interplay between the constraints of the population and the exploratory drive of individual cells. At one end of the spectrum, normal tissues are characterized by high tissue constraints and low exploration. Tissue homeostasis is maintained by various mechanisms that include biochemical and mechanical interactions between cells. At the other end of the spectrum, cancer progression is associated with looser tissue constraints that allow a high exploratory behavior. This view shares with the Tissue Organization Field Theory (TOFT) of cancer the premise that cells are not quiescent by default, but rather maintain continuous internal drive, while homeostasis is enforced on them by the tissue level of organization (Sonnenschein and Soto, 2020). It is also consistent with the idea that aging affects the propensity for cancer development through the decrease in effective tissue homeostasis (Capp and Thomas, 2021). Our conceptual framework may open the door to novel directions in cancer research and therapeutic development.

**Acknowledgements**
This work was supported in part by the Israeli Science Foundation (Grant no. 346/16, O.B.; and Grant No. 155/18, N.B.). We acknowledge the Adams Fellowship Program of the Israel Academy of Science and Humanities (AS).